\documentclass{ws-procs10x7}

\newcommand{\ra}{\mbox{$\rightarrow$}}
\newcommand{\Ra}{\mbox{$\Rightarrow$}}

\newcommand{\zpr}{\mbox{$Z'$}}

\newcommand{\uprm}{\mbox{$U(1)'$}}

\newcommand{\skipblk}[1]{}                                                      
\def\bqa{\begin{eqnarray}}                                                      
\def\eqa{\end{eqnarray}}

\newcommand{\x}{\mbox{$\times$}}

\newcommand{\beq}{\begin{equation}}                                             
\newcommand{\eeq}{\end{equation}}

\def\mxth{\mathsurround=0pt }
\def\xversim#1#2{\lower2.pt\vbox{\baselineskip0pt \lineskip-.5pt
  \ialign{$\mxth#1\hfil##\hfil$\crcr#2\crcr\sim\crcr}}}             
\def\simgr{\mathrel{\mathpalette\xversim >}}                                    
\def\simle{\mathrel{\mathpalette\xversim <}}  
\begin{document}

\title{Neutrino Physics (theory)}

\author{Paul Langacker}

\address{Department of Physics, University of Pennsylvania, Philadelphia, PA 19104,
USA\\E-mail: pgl@hep.upenn.edu}

\twocolumn[\maketitle\abstract{Nonzero neutrino masses are the first definitive need 
to extend the standard model. After reviewing the basic framework, I describe the status of some
of the major issues, including tests of the basic framework of neutrino masses and mixings;
the question of Majorana vs. Dirac; the spectrum, mixings, and number of neutrinos;
models, with special emphasis on constraints from typical superstring constructions (which
are not consistent with popular bottom-up assumptions); and other implications.
}]

\section{Neutrino Preliminaries}

\subsection{Definitions}\label{definitions}
A {\bf Weyl fermion} is the minimal fermionic field.
It has two degrees of freedom of opposite chirality, related by
CPT or Hermitian conjugation, such as
 $\psi_L \leftrightarrow \psi^c_R$. (Which is called the particle and which
 the antiparticle is a matter of convenience.)
An {\bf active}  (a.k.a. \textbf{ordinary} or \textbf{doublet}) neutrino $\nu_L$ is in an
 $SU(2)$ doublet with a charged lepton partner and therefore has
  normal weak interactions. Its CPT conjugate
  $ \nu^c_R$ is a right-handed antineutrino.
  A {\bf sterile}  (a.k.a. \textbf{singlet} or \textbf{right-handed})  
  neutrino $N_R \leftrightarrow N^c_L$ is an
 $SU(2)$ singlet. It has no interactions except by mixing, Higgs couplings, or
beyond the standard model (BSM) interactions. 
 Sterile neutrinos are present in almost
 all extensions of the standard model. The only questions are whether they are light and
 whether they mix with the active neutrinos, as suggested by the LSND experiment.
 
Fermion mass terms convert a spinor of one chirality into the other. A {\bf Dirac
mass} connects two distinct Weyl spinors (usually active to sterile), such as
 $m_D \bar{\nu}_L N_R + h.c.$ There are four components ($\nu_L, N_R$ and
their conjugates), and one can define a conserved lepton number $L$.
An active-sterile Dirac mass violates weak isospin by 1/2 unit, $\Delta I=\frac 1 2$, and can be
generated by a Yukawa coupling to a Higgs doublet. This is analogous to quark and
charged lepton masses, but raises the question of why $m_D$ (i.e., the Yukawa coupling)
is so small. 
There are variant types of Dirac masses  in which an active $\nu_L$ is coupled
to a different flavor of  active $\nu^c_R$, e.g., $m_D \bar{\nu}_{eL}
  \nu^c_{\mu R}$, in which $ L_e-L_\mu$ conserved. This has $\Delta I=1$
  and may emerge as a limit of a model with Majorana masses.
  
  \begin{figure}
     \setlength{\unitlength}{0.9mm}
     \begin{picture}(65,38)
     \thinlines
\thicklines
\put(15,0){\vector(0,1){6}}
\put(15,6){\line(0,1){18}}
\put(15,24){\vector(0,1){10}}
\put(15,34){\line(0,1){4}}
\multiput( 15,19)(4,0){4}{\line(1,0){2}}
\put(33,19){\circle{6.5}}
\put(5,29){$\nu_L$}
\put(7,17){$h$}
\put(5,5){$N_R$}
\put(25,27){$v=\langle \phi \rangle$}
\put(24,7){$m_D = h v$}
\put(55,0){\vector(0,1){6}}
\put(55,6){\line(0,1){18}}
\put(55,24){\vector(0,1){10}}
\put(55,34){\line(0,1){4}}
\put(55,19){\line(1,1){4}}
\put(55,19){\line(1,-1){4}}
\put(55,19){\line(-1,1){4}}
\put(55,19){\line(-1,-1){4}}
\put(60,29){$\nu_L$}
\put(60,5){$\nu^c_R$}

\end{picture} 
\caption{Dirac and Majorana  masses.}
\label{fig:masses}
\end{figure}
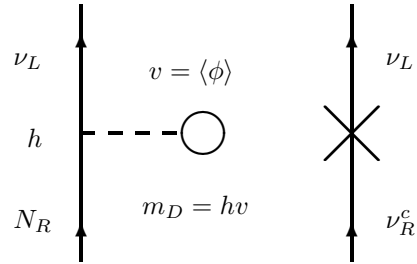

A {\bf   Majorana Mass} 
connects a Weyl spinor with its own CPT conjugate.
There are only two components, and lepton number is 
necessarily violated by two units, $\Delta L=\pm 2$.
An active neutrino Majorana mass term
     $\frac 1 2 (m_T \bar{\nu}_L \nu^c_R + h.c.)$
     has $\Delta I = 1 $,
     requiring a Higgs triplet or a higher dimensional operator with two
     Higgs doublets. A sterile Majorana mass term
 $\frac 1 2 (m_S \bar{N}^c_L N_R + h.c.)$ has $\Delta I = 0$. It could in principle
 be a bare mass, but most concrete models have additional constraints and require
 the expectation value of a Higgs singlet.

One can also  consider mixed models involving both Majorana and Dirac mass terms.
For example, the case $m_S$ (or $m_T$) $ \ll m_D$ involves two almost degenerate
Majorana neutrinos (pseudo-Dirac\footnote{A Dirac neutrino can be thought of as the
limiting case, with two degenerate Majorana neutrinos with maximal mixing and opposite
CP parity.}). Another
 well-known example is the seesaw model, in which  $m_S \gg m_D$
and $m_T=0$. 

For three families the most general ($6\x 6$) mass matrix is
\beq  L = \frac{1}{2} \left( \bar{\nu}_L \; \; \bar{N}_L^c \right) \left(
\begin{array}{cc} {m}_{T} & {m}_{D} \\ {m}_{D}^{T} &
{m}_{S} \end{array} \right) \left( \begin{array}{c} \nu_R^c \\ N_R
\end{array} \right) + {\rm  hc}, \label{three} \eeq
where  $\nu_L$ ($N_R$) represent 3 flavors  of active (sterile) neutrinos,
and the $3\x 3$ submatrices are (a) the active Majorana mass 
matrix\footnote{The off diagonal terms in $m_{T,S}$ are still considered Majorana 
as long as there is no way to define a conserved $L$.}
$m_T=m_T^T$, generated by a Higgs triplet; (b) 
 the Dirac mass matrix $m_D$, generated by a Higgs doublet; and (c)
 the sterile Majorana mass matrix  $m_S=m_S^T$, generated by a SM singlet.

\subsection{Neutrino Mass Patterns}
The Solar neutrino oscillation parameters~\cite{reviews} are now confirmed by SNO and Kamland
to fall in the large mixing angle (LMA) region, with 
$\Delta m^2_\odot \sim 8 \x 10^{-5}$ eV$^2$ and large but nonmaximal mixing angle $\theta_\odot$.
The atmospheric neutrinos are characterized by 
$|\Delta m^2_{\rm  Atm}| \sim 2 \x 10^{-3}$ eV$^2$,
with large mixing consistent with maximal. Reactor experiments establish that
the third angle, $U _{e3}$ is small. For maximal atmospheric mixing and neglecting
$U _{e3}$ this would imply mass eigenstates
 \begin{eqnarray} 
  \nu_3 & \sim & \nu_+  \nonumber \\
  \nu_2 & \sim & \cos \theta_\odot \ \nu_- - \sin \theta_\odot \ \nu_e  \\
  \nu_1 & \sim & \sin \theta_\odot \ \nu_- + \cos \theta_\odot \ \nu_e,  \nonumber
   \end{eqnarray} 
where $\nu_\pm \equiv \frac{1}{\sqrt{2}} \left( \nu_\mu \pm \nu_\tau \right)$.
Depending on the sign of $\Delta m^2_{\rm  Atm}$ one can have either the
normal hierarchy, which is analogous to the quarks and charged leptons, or the inverted 
one. 
Degenerate patterns refer to the possibility that the overall mass scale,
which is not determined by oscillation experiments,
is large compared to the mass 
differences. This was once strongly motivated by hot dark matter (HDM) scenarios,
but now HDM serves only as an upper limit. 
The inverted and degenerate patterns may be radiatively unstable~\cite{radiative}.

  \begin{figure}
     \setlength{\unitlength}{0.9mm}
     \begin{picture}(65,38)
     \thinlines
\thicklines
\put(10,4){\line(1,0){23}}
\put(10,10){\line(1,0){23}}
\put(10,29){\line(1,0){23}}
\put(3,3){1}
\put(3,10){2}
\put(3,28){3}
\put(45,4){\line(1,0){23}}
\put(45,23){\line(1,0){23}}
\put(45,29){\line(1,0){23}}
\put(38,3){3}
\put(38,20){1}
\put(38,27){2}

\end{picture} 
\caption{The normal and inverted hierarchies.}
\label{patterns}
\end{figure}
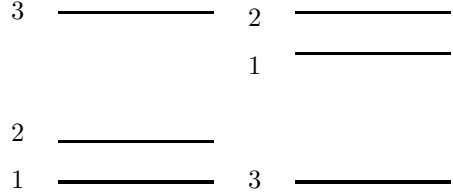

\section{The Basic Framework}
Many ideas~\cite{reviews} have previously
been put forward as alternatives to oscillations amongst the 3 active neutrinos
for the Solar and atmospheric neutrinos.
These include oscillations into sterile neutrinos; neutrino decay; decoherence
between the quantum components of the wave function;
new flavor changing interactions;
Lorentz, CPT, or equivalence principle violation~\cite{lorentz}; and (for the Solar neutrinos)
large magnetic moments and resonant spin flavor precession (although there are
strong constraints from stellar cooling)~\cite{rsfp}. 
These could typically describe the contained (lower energy)
atmospheric events, but most schemes were excluded by (higher energy) upward throughgoing
events, e.g., because they  depend on $LE$ or $L$ rather than $L/E$. This conclusion
has been further strenghtened by observation of a dip in the $L/E$ spectrum by
SuperKamiokande, a clear indication of oscillations.
Similarly, the KamLAND reactor results eliminated alternatives to the LMA oscillation
picture for the Solar neutrinos; this has also been strenghtened by a suggestion of a
dip in the KamLAND spectrum. The data is also now sufficiently good
to indirectly demonstrate the need for MSW (matter) effects in the Sun~\cite{mswevidence},
although direct confirmation would require observation of the transition between
the low (high) energy vacuum (MSW) regimes in a future $pp$ neutrino experiment.

Thus, these alternatives are now excluded as the dominant mechanism.
Emphasis has shifted to precision theoretical~\cite{solartheory}
and experimental
studies to search for or constrain such effects as small perturbations\footnote{For example,
small flavor changing operators~\cite{fcnc1} could shift the Solar parameters slightly, to coincide
with a different KamLAND oscillation minimum~\cite{fcnc}.}
 and to further
test the Standard Solar Model.

\section{Majorana or Dirac}
 One of the most important questions for understanding the origin of neutrino masses is whether they
 are Majorana or Dirac. Many theorists are convinced that they { must} be Majorana, because
 (a) no standard model gauge symmetry forbids Majorana masses; (b) 
 nonperturbative electroweak processes (sphalerons) and black holes violate $L$
 (in practice, such effects are negligibly small); and (c)
 standard grand unified theories (GUTS) violate $L$.
 However, these arguments are not compelling because
 there could be additional symmetries to forbid or strongly suppress $L$ violation,
analogous to the strong suppression of proton decay. For example, there could well be
 new gauge symmetries (e.g., a   \zpr~\cite{zprime}) at the TeV-scale which could forbid Majorana 
 masses. Similarly, constraints in superstring
 constructions are extremely restrictive and could forbid or suppress 
 them~\cite{string}.
Therefore,   Dirac or pseudo-Dirac masses are serious possibilities. 

The only practical  way to distinguish Majorana and Dirac masses experimentally is neutrinoless
double beta decay ($\beta \beta_{0\nu}$)~\cite{bb}. If observed, $\beta \beta_{0\nu}$ would imply
Majorana masses\footnote{$\beta \beta_{0\nu}$ could also be driven by new interactions,
such as $R$-parity violation in supersymmetry. These would also lead to Majorana masses
at some level.}. As discussed below, the converse is {\em not} true.

\section{The Spectrum}
Another key uncertainty (and constraint on models)
is whether there is a normal, inverted, or degenerate spectrum.
It should eventually be possible to distinguish the normal and inverted hierarchies using
long baseline oscillation effects, because the MSW matter effects associated with 
$\Delta m^2_{\rm  Atm}$ change sign. It may also be possible to distinguish from
the observed energy spectrum in a future supernova because of matter effects in the
supernova and in the Earth~\cite{supernova}. Planned and proposed $\beta \beta_{0\nu}$
experiments would be sensitive to Majorana masses predicted by the
inverted and degenerate spectra. Unfortunately, nonobservation could be due either
to a normal hierarchy or to Dirac masses.

There are three complementary future probes of the absolute mass scale:
\begin{itemize}
\item  Tritium beta decay experiments measure the quantity, $m_\beta \equiv \Sigma_i |U_{ei}|^2 |m_i|$,
where $U$ is the leptonic mixing matrix. The KATRIN experiment should  be sensitive to
$m_\beta \sim 0.2$ eV, compared to the present upper limit of 2 eV.
It could only see a signal for the degenerate cases. 
\item Cosmological 
(large scale structure) observations are sensitive to $\Sigma \equiv   \Sigma_i |m_i|$.
The most stringent
claimed limit is $\Sigma \simle 0.42$ eV~\cite{cosmological}. 
Using future Planck data, it may be possible
to extend the sensitivity down to $0.05-0.1$ eV, close to the minimum value
 0.05 eV $\sim 
\sqrt{|\Delta m^2_{\rm  Atm}|}$
allowed by the oscillation data. However, there are 
significant theoretical uncertainties.
\item The $\beta \beta_{0\nu}$ amplitude is proportional to the effective mass
$m_{\beta \beta} \equiv\Sigma_i U_{ei}^2 m_i $, where there
can be cancellations due to signs or (Majorana) phases in
$U_{ei}^2$ or in $ m_i$ (depending on conventions). Proposed experiments
would be sensitive to $m_{\beta \beta} \sim 0.02$ eV, corresponding to
 Majorana masses predicted by the inverted or degenerate
spectra. If observed there would be a significant uncertainty in the
actual value of $m_{\beta \beta}$ due to the  nuclear matrix elements~\cite{matrixelements}.

There is a claimed observation~\cite{germanium} of $\beta \beta_{0\nu}$, corresponding to
$0.17 <  m_{\beta \beta} < 2.0$ eV. This would be extremely important if confirmed,
but would not be easy to reconcile with the current expectations from cosmology and
oscillations, as shown in Figure \ref{massscale} (from~\cite{scale}).
 \begin{figure}
\epsfxsize160pt  
\figurebox{160pt}{160pt}{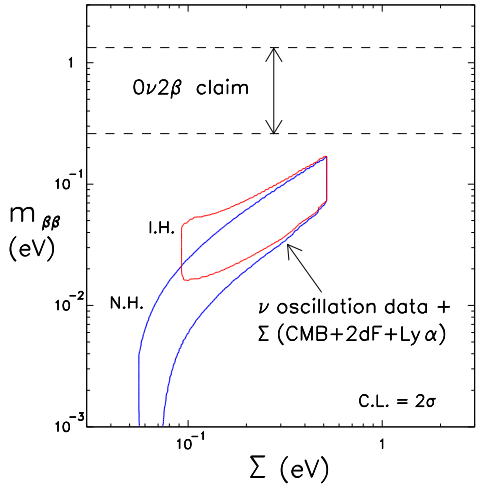}  
\vspace{-1cm}
\caption{Cosmological and oscillation constraints on $\Sigma$ and
$m_{\beta \beta}$ for the normal (N.H.) and inverted (I.H.) hierarchies.}
\label{massscale}
\end{figure}
\end{itemize}

\section{Neutrino Mixings}
\label{numixings}
The leptonic mixing (Pontecorvo, Maki, Nakagawa, Sakata~\cite{bilenky}) matrix $U_{PMNS} $
is due to the mismatch of the charged lepton and neutrino mixings, 
 $U_{PMNS} = U_e^\dagger U_\nu\equiv U.$ 
It is very different from the quark mixing matrix $U_{CKM}$. Whereas the latter has small mixing
angles, two of the leptonic mixings are large. The atmospheric angle
$\theta_{23}$ is
consistent with maximal, while the 
Solar  angle is large but not maximal, $\tan^2 \theta_{12} =0.40^{+0.09}_{-0.07}$.
On the other hand, the third angle 
 $\sin^2 \theta_{13} < 0.03$ (90\%). A better knowledge is important because the angles
 may be a critical test of models, especially the deviation of $\theta_{12}$ from maximal
 and the value of $\theta_{13}$. (The latter is also of some urgency because leptonic
 CP violation in oscillations vanishes for $\theta_{13}=0$.)
 
 The observed large mixings came as something of a surprise, especially in frameworks
 such as grand unification, in which the simplest models would yield small
 mixings similar to those in  $U_{CKM}$. The mixings can be associated with either
 the charged leptons ($U_e$) or the neutrinos ($U_\nu$), or both. 
 One can always choose a basis in the space of lepton families in which one or the
 other is the identity matrix, but that basis might not be the one in which the family or
 other symmetries or constraints are most apparent. Until recently, most
 models assumed either
$U_e\sim I$ or $U_\nu \sim I$, with the large mixings due to the other sector. In this case
it is easier to achieve bimaximal mixing, $\theta_{23}=\theta_{12}=\frac{\pi}{4}$, than
 the observed $\theta_{12}$. Recently, several authors have pointed out that the
 observations are consistent with a bimaximal $U_\nu$ and small (Cabibbo-like) deviations
due to $U_e\ne I$. (In fact,  the central value of $\theta_{12}$ is $\sim \frac{\pi}{4}  -\theta_{\rm  Cabibbo}$.)~\cite{charged}.

\section{The Number of Neutrinos}
There are two major constraints on the number of neutrino types.
The invisible $Z$ width implies $N_\nu = 2.9841(83)$, where $N_\nu$ is the
number of active  neutrinos with $m_\nu < M_Z/2$. Clearly, there is room for only three.
Other unobserved new particles
from $Z$ decay would also give a positive contribution to $N_\nu$. 

A complementary constraint comes from big bang nucleosynthesis (BBN)~\cite{BBN}, 
in which the predicted $^4He$
abundance depends sensitively on the competition between weak 
and expansion rates, and therefore on the number of relativistic particles present
at $T\sim $1 MeV. This implies $N_\nu' < 3.1-3.3$, where  $N_\nu' $ counts
the active $\nu$'s with $m_\nu \simle 1$ MeV. It also includes sterile 
$\nu$'s, which could be produced by oscillation effects, for a wide range of masses and mixings
with the active neutrinos~\cite{bbnsterile,bbns2}. It does  {\em not} include $N_R$ for light Dirac 
neutrinos unless they could be produced by new BSM interactions~\cite{zpbbn}.

\subsection{LSND}
LSND has claimed evidence for oscillations, especially $\bar{\nu}_\mu \ra \bar{\nu}_e$,
with $|\Delta m^2_{\rm  LSND}| \simgr 1$ eV$^2$. If this is confirmed by the Fermilab
MiniBooNE experiment, it would strongly suggest the existence of one or more light
sterile neutrinos which mix with the active $\nu$'s of the same chirality
(since the $Z$ width does not
allow a fourth active neutrino)\footnote{An alternative, nonstandard neutrino
interactions~\cite{pakvasa}, is strongly disfavored by the KARMEN data.}.
 Such sterile neutrinos are present in
most models, but the active-sterile mixing would require Dirac {\em and} Majorana mass terms
which are both tiny and either comparable
or $m_S \ll m_D$ (pseudo-Dirac), which are difficult to achieve theoretically\footnote{Some possible
mechanisms and described in~\cite{sterile1,sterile2}.
For a review, see~\cite{bbns2}}.
It is also difficult to accomodate experimentally. It is well established that neither the 
Solar nor the atmospheric oscillations are {\em predominantly} into sterile states.
Furthermore, a combination of Solar, atmospheric, Kamland, and reactor and
accelerator disappearance limits exclude both the $2+2$ and $3+1$ schemes~\cite{fournu}.
These refer respectively to the possibilities that the Solar and atmospheric pairs
(which contain admixtures of sterile) are separated by a gap of $\sim $ 1 eV,
and to the possibility of three closely spaced (mainly active) states separated from
the fourth (mainly sterile) state by $\sim$ 1 eV.
However, some 5 $\nu$ (i.e., $3+2$) patterns involving mass splittings around 1 eV$^2$ and
20 eV$^2$ are more successful~\cite{fivenu}.
\begin{figure}
         \setlength{\unitlength}{0.9mm}
     \begin{picture}(45,45)(0,10)
\thicklines
\put(5,15){\line(1,0){25}}
\put(5,18){\line(1,0){25}}
\put(5,39){\line(1,0){25}}
\put(5,45){\line(1,0){25}}
\put(45,15){\line(1,0){25}}
\put(45,18){\line(1,0){25}}
\put(45,21){\line(1,0){25}}
\put(45,45){\line(1,0){25}}
\end{picture} 
\caption{$2+2$ and $3+1$ patterns.}
\label{fig:fournu}
\end{figure}
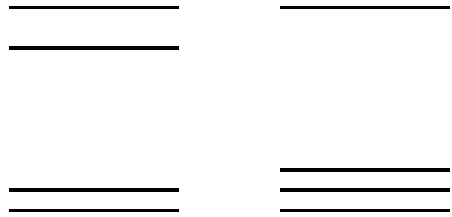

The constraints on sterile neutrinos
from BBN and large scale structure are also severe. The $2+2$ and $3+1$ 
patterns\footnote{The $3+2$ schemes have not been
studied in detail, but are also likely to be problematic.}
 are again 
apparently excluded~\cite{bbnsterile,bbns2}, 
although there are some (highly speculative/creative)
loopholes. 
These include:
(a) Large $\nu$ asymmetries to suppress or compensate the steriles~\cite{bbnasym}.
(b) Late time phase transitions to suppress $\nu$ masses and sterile mixings until
after neutrino decoupling ($T <$ 1 MeV)~\cite{late}.
(c) Time varying $\nu$ masses (due to coupling to special scalar fields) so that the
steriles were too massive to produce cosmologically (with possible implications
for matter effects and for dark energy)~\cite{timevarying}.
(d) A low reheating temperature after inflation~\cite{reheat}.
On the other hand, sterile neutrinos could play a useful role in $r$-process
nucleosynthesis~\cite{rprocess} and in understanding pulsar kicks~\cite{pulsar}.

Instead of mixing with sterile neutrinos, one can invoke CPT violation, which could manifest
itself as a difference in the neutrino and antineutrino masses, allowing three mass
differences for active neutrinos and antineutrinos. The original proposal~\cite{cpt1}
was excluded by Kamland, which observed $\bar{\nu}_e$ disappearance for
the Solar parameters. An alternative, which mainly affects the atmospheric oscillations~\cite{cpt2},
has been shown to be excluded by  global data analyses~\cite{fournu}.
Still surviving is a hybrid scheme which also invokes a sterile neutrino~\cite{cptsterile}.
Another possibility is that CPT violation manifests itself in quantum decoherence
rather than in the masses~\cite{decoherence}.

\section{Models of Neutrino Mass}
There are an enormous number of models of neutrino mass~\cite{reviews}.
Models constructed to yield small Majorana masses include:
the ordinary (type I) seesaw;
models with heavy Higgs triplets (type II seesaw);
TeV (extended) seesaws, with $m_\nu \sim m^{p+1}/M^p$, e.g., with $M$ in the TeV range;
radiative masses (i.e., generated by loops);
supersymmetry with $R$-parity violation;
mass generation by terms in the K\" ahler potential;
anarchy (random entries in the mass matrices);
large extra dimensions (LED),  possibly combined with one of the above. 

Small Dirac masses may be due to:
 higher dimensional
 operators (HDO) in intermediate scale models (e.g., associated with a \uprm \ or supersymmetry breaking); large intersection areas in intersecting brane models~\cite{braneareas}; or
large extra dimensions, from volume suppression if $N_R$ propagates in the bulk~\cite{largedim}.

I will only describe a few of these in more detail, as well as comment on the possibilities
in superstring constructions, which may lead to variant forms~\cite{string}.

\subsection{Textures}
Neutrino textures~\cite{textures,low} are specific guesses about form of the 
$3 \x 3$ neutrino mass matrix
or the Dirac and Majorana matrices entering seesaw models. These are often studied 
in connection with models also involving quark
and charged lepton mass matrices, 
such as grand unification (GUTs), family symmetries, or left-right symmetry.
These textures are not unique, even given perfect information about the quark and
lepton masses and mixings, because the forms are changed when one rotates to 
different bases for the left and right-handed fermions. However, it is hoped that
there is some basis in which the underlying symmetries of the theory are especially
simple, and that finding a successful texture (typically involving hierarchies of large
and small elements) will provide a clue. Some examples, which lead to each of the
possible neutrino hierarchies, are:
\bqa
  {\rm  Normal:} && \  m_\nu=  m \left(
\begin{array}{ccc} \epsilon^2& \epsilon &  \epsilon  \\  \epsilon  & 1 & 1 \\
 \epsilon & 1 & 1\end{array} \right), \ \epsilon \ll 1   \nonumber \\
 {\rm Inverted:}  && \  m_\nu=  m \left(
\begin{array}{ccc} 0 & 1 &  1 \\  1  & 0 & 0 \\
1 & 0 & 0\end{array} \right) + \ {\rm small}\nonumber  \\
  {\rm  Degenerate:}  &&  \  m_\nu=  m I  + \ {\rm  small}.  
\eqa
Another complication is that most models predict the form of the textures at the
Planck or GUT scales,  so that one must run the mass and
mixing parameters to low energy to compare with experiment.
For the neutrinos this is especially important for degenerate or inverted cases
$m_i\sim m_j$ (with the same sign), in which case there may either be instabilities
or the radiative generation of large mixings~\cite{radiative}.

\subsection{Dirac Masses}
\label{diracmass}
One promising mechanism for small Dirac neutrino (or other) masses
is that elementary Yukawa couplings $L N^c_L H_2$ may be forbidden by new 
symmetries (e.g., \uprm) of the low energy theory or
      by string constraints, but that very small effective couplings
       are generated by higher dimensional operators, such as
\bqa L_\nu &\sim& \left( \frac{S}{M_{Pl}}\right)^p L N^c_L H_2, \  \  \  
     \langle S \rangle \ll M_{Pl} \nonumber \\
&\Rightarrow& m_D \sim \left( \frac{\langle S \rangle}{M_{Pl}}\right)^p  
      \langle H_2 \rangle. \eqa
      $S$ is a standard model singlet field assumed to acquire a VeV.
        For large $p$ one could have $\langle S\rangle$ close to $M_{Pl}$,
        as can occur in heterotic string models with an anomalous \uprm~\cite{anomalous}.
For smaller $p$, $\langle S\rangle$ could be at an intermediate scale $\ll M_{Pl}$,
e.g., $\langle S\rangle \sim 10^8$ GeV for $p=1$. The scale of $\langle S\rangle$
could be associated with the breaking of a non-anomalous \uprm \ along a $D$ and (almost) $F$ flat direction~\cite{sterile2}, or with supersymmetry breaking~\cite{susyint}.
A variant on such models could have additional operators that naturally yield
ordinary-sterile mixing should it be needed. Such mechanisms are compatible with
the general features of string constructions, but there have been no detailed models.

\subsection{The (Ordinary) Seesaw Model}
The ordinary (type I) seesaw~\cite{seesaw} is the most popular model for small 
neutrino masses. It assumes that
$m_T=0$ in (\ref{three}) and that the eigenvalues of  $m_S$ are $\gg m_D$ (e.g., $10 ^{12}$ GeV),
yielding an effective Majorana mass matrix
\beq m_\nu^{\rm  eff} = - m_D m_S^{-1} m_D^T \eeq
for the 3 light active $\nu$'s.

The ordinary seesaw is usually implemented in connection with grand unification.
For example, in $SO(10)$ the $N_R$ occurs naturally, although large Higgs multiplets
such as the 126 (or higher dimensional operators) must be invoked to generate
$m_S$.
Most of the explicit $SO(10)$ models yield the normal hierarchy~\cite{albright}.

  \begin{figure}
\setlength{\unitlength}{0.9mm} 
\small
\begin{picture}(65,30)(4,10)
\thicklines
\put(29,25){\line(-1,1){14}}
\multiput(29,25)(-7,-7){2}{\line(-1,-1){5}}
\put(29,25){\line(1,0){25}}  
\put(54,25){\line(1,1){14}}  
\multiput(54,25)(7,-7){2}{\line(1,-1){5}}
\put(11,38){$\nu$} 
\put(70,38){$\nu$}
\put(40,28){$N$}
\put(11,10){$H_2$}
\put(68,10){$H_2$}
\end{picture} 
\caption{The ordinary seesaw. $N$ is a heavy Majorana neutrino.}
\label{fig:seesaw}
\end{figure}
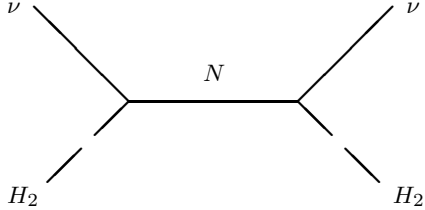


The grand unified theory seesaw model~\cite{seesaw} is an elegant
mechanism for generating small Majorana neutrino masses, which leads
fairly easily to masses in the correct range.  It  also provides a
simple framework for leptogenesis~\cite{leptogenesis}, in which the
decays of heavy Majorana neutrinos produce a lepton asymmetry, which is
later partially converted to a baryon asymmetry by electroweak sphaleron
effects.

However, the expectation of the simplest grand unified theories is
that the quark and lepton mixings should be comparable and that the
neutrino mixings 
should be small, rather than the large mixings that
are observed. This can be evaded in more complicated GUTs, e.g.,
(a) those involving   highly non-symmetric (lopsided)
mass matrices~\cite{reviews}, in which
there are large mixings in the right-handed charge $-1/3$ quarks
(where it is unobservable) and in the left-handed charged leptons $U_e$
(this is harder to achieve in $SO(10)$ than $SU(5)$);
or (b) in those with complicated textures~\cite{textures} for the heavy
Majorana neutrino mass matrix, i.e., involving an $m_D, m_S$ conspiracy to give large $U_\nu$ 
mixings.
However,  the need to do so makes the GUT
seesaw concept less compelling. Furthermore, a number of promising
extensions of the standard model or MSSM do not allow the canonical
GUT seesaw. For example, the large Majorana masses needed are often
forbidden, e.g., by extra \uprm \ symmetries~\cite{zprime} predicted in many string
constructions. Similarly, it is difficult to accomodate
traditional grand unification (especially the needed adjoint and high
dimension Higgs multiplets needed for GUT breaking and the seesaw)
in simple string constructions. 
Such constructions also
tend to forbid direct Majorana mass terms and large scales.  
Finally, the active-sterile neutrino mixing
required in the
schemes motivated by
the LSND experiment is difficult to implement in canonical
seesaw schemes.

\subsection{Triplet models}

An alternative class of models involves the introduction of a
 Higgs triplet $T=(T ^{++} \ T^+ \ T^0)^T$ with weak hypercharge $Y=1$.
Majorana masses $m_T$ can then be generated from 
the Yukawa couplings $L_{\nu}= {\lambda}^{T}_{ij} L_i T L_j$
if $\langle T^0 \rangle\ne 0$ but $\ll$ the electroweak scale.
An early version, the  Gelmini-Roncadelli model~\cite{GRmodel},
assumed spontaneous $L$ violation.
The original model has been excluded because the decay 
$Z \ra$ scalar $+$ Majoron (the Goldstone boson of $L$ violation)
would increase the $Z$ width by the 
equivalent to two extra neutrinos. This can be evaded in
invisible Majoron models~\cite{reviews}, in which the Majoron is mostly singlet.

However, most of the more recent
 triplet models~\cite{triplet} assume a very heavy triplet mass $M_T$, and
break $L$ explicitly by including $THH$ couplings,
giving large a Majoron mass.
This coupling will induce a very small 
VeV for the triplet suppressed by $M_T^{-1}$ (the type II seesaw).
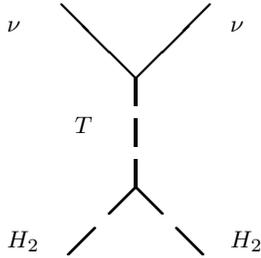
\begin{figure}[htbp]
\setlength{\unitlength}{0.9mm} 
\small
\begin{picture}(60,40)(-5,12)
\thicklines
\put(29,43){\line(-1,1){11}}
\put(29,43){\line(1,1){11}}
\multiput(29,27)(0,6){3}{\line(0,1){4}}
\multiput(29,27)(-6,-6){2}{\line(-1,-1){4}}
\multiput(29,27)(6,-6){2}{\line(1,-1){4}}
\put(10,50){$\nu$}
\put(43,50){$\nu$}
\put(20,35){$T$}
\put(10,18){$H_2$}
\put(43,18){$H_2$}
\end{picture} 
\normalsize
\caption{The triplet seesaw}
\label{trip}
\end{figure}
The type  II models are often considered in the context of $SO(10)$ or
left-right symmetry, with both the ordinary and
triplet mechanisms competing and with related parameters. However, they can also
be considered independently.
A general supersymmetric version would involve the superpotential
\begin{eqnarray} 
W_{\nu} &=& {\lambda}^{T}_{ij} L_i T L_j + \lambda_1 H_1 T
H_1 + \lambda_2 H_2 \bar{T} H_2  \nonumber \\
 & & + M_T T \bar{T} + \mu H_1 H_2 , \label{tripletw}
 \end{eqnarray}
where $T,\ \bar{T}$ are triplets with $Y=\pm1$,  and typically
$M_T \sim 10 ^{12}-10 ^{14}$ GeV.
This induces a seesaw-type VeV and triplet Majorana mass
\beq  \langle T^0 \rangle \sim -\lambda \frac{ \langle H_2^0\rangle^2}{M_T}\ \ \ \Ra \ \
  {m}_{\nu ij} =  -\lambda^T_{ij}\lambda_2
\frac{v_2^2}{M_T}. \eeq
Equivalently, one can integrate out the heavy triplet, inducing a higher dimensional
effective mass operator coupling two lepton doublets to two Higgs doublets,
$ \frac{1}{2M}{\lambda}^{\nu}_{ij}\left(L_i H_2\right)\left(L_j H_2\right)$. 

\section{Neutrinos in String Constructions}
Some of the key ingredients of most GUT and bottom up models are either absent or
different in known semi-realistic string constructions,
both heterotic and intersecting brane~\cite{string}.
In particular, string constructions typically yield
bifundamental, singlet, and adjoint representations of the gauge groups, not the
   large representations  usually invoked in GUT model building. Moreover,
string symmetries and constraints may forbid couplings allowed by the apparent symmetries
of the four-dimensional field theory. In particular, 
the superpotential terms leading to Majorana masses may be absent, 
or if they are present they may not be diagonal (i.e., connecting the same family or same 
flavor),
leading to nonstandard mass matrices.
Furthermore,  GUT Yukawa relations are typically broken.
Another difference is that  the nonzero
 superpotential terms are related to gauge couplings, so they may naturally be equal or simply
 related, with hierarchies in effective Yukawas typically due to higher-dimensional operators
 in heterotic models or due to the areas of intersection triangles in intersecting brane constructions.

\subsection{The Seesaw in String Constructions} 
There seem to be no fundamental difficulties in generating Dirac masses in string constructions,
which can be small by the mechanism described in Section \ref{diracmass}. However, 
Majorana masses are more difficult.
Several questions arise when one attempts to embed the seesaw model:
(a) Can one generate a large effective $m_S$ from superpotential terms like
\beq W_\nu \sim c _{ij}  \frac{S^{q+1}}{M_{Pl}^q} N_i N_j \  \  \  
       \Rightarrow (m_S) _{ij} \sim c _{ij}  \frac{\langle S \rangle^{q+1}}{M_{Pl}^q},   \eeq
       consistent with $D$ and $F$ flatness? ($S^{q+1}$ can represent a product of
       $q+1$ different SM-singlet fields, or can even contain SM-nonsinglets if one allows
       for condensations of products of fields.)
(b) Can one have such terms simultaneously with Dirac couplings, consistent
with flatness and other constraints? If (a) and (b) are satisfied, there is a possibility
of an ordinary seesaw. However, to obtain anything resembling typical bottom-up constructions,
there are two more conditions.
(c) Is $c _{ii}=0$? This is motivated by the fact that diagonal terms in the superpotential are
rare in string constructions. One could still have a seesaw with off diagonal terms, but
it would be very nonstandard. Finally (d):
Are  the assumptions usually made in bottom-up constructions 
relating the neutrino sector  to the quark and charged lepton masses 
maintained?

There have been relatively few detailed investigations 
of neutrino masses in string models, but none have been consistent with all
of these assumptions. Some constructions lead to a conserved $L$ and Dirac masses~\cite{stringl}.
Only one construction~\cite{flipped}, based on flipped $SU(5)$, has found a possibly flat direction
that can yield an ordinary seesaw, but that one is very non GUT-like in detail.
A detailed study of  the $Z_3$ orbifold is in progress~\cite{string}, 
but so far has yielded no Majorana mass terms.

Another possibility is an extended (TeV-scale) seesaw~\cite{extended}, 
in which the light neutrino masses 
are of order $m_\nu \sim m^{p+1}/m_S^p$, with $p>1$ (e.g., $m\sim 100$ MeV, $m_S\sim 1$ TeV
for $p=2$). This could come about, for example, by the mass matrix
\beq \frac{1}{2} \left( \bar{\nu}_L \; \; \bar{N}_L^c  \; \; \bar{N'}_L^c \right) \left(
\begin{array}{ccc} 0 & {m}_{D} &0 \\
 {m}_{D}^{T} & 0 & {m}_{SS'} \\
0& {m}_{SS'}^T  & m_{S'}
\end{array} \right) \left( \begin{array}{c} \nu_R^c \\ N_R \\ N'_R
\end{array} \right) \eeq 
where $\nu_L$, $N_R$, and $N'_R$  each represent three flavors, ${m}_{SS'}$
has TeV-scale eigenvalues, and $m_D$ and $m_{S'}$ are much smaller.
This may occur in certain heterotic constructions
 (depending on dynamical assumptions)~\cite{stringe}.
 
\subsection{Triplets in String Constructions}
If the triplet model in (\ref{tripletw}) were embedded in a string construction then one
expects $\lambda^T_{ij}\sim 0$ for $i=j$, i.e., that the diagonal terms vanish
at the renormalizable level, implying that $ {m}_{\nu ii}=0$ to leading order\footnote{One
would also need either to have multiple  Higgs doublets $H _{1,2}$ with $ \lambda _{1,2}$ off
diagonal or to generate the $ \lambda _{1,2}$ by HDO. It would only
be necessary for one pair to survive to low energies.}.
That is because the existence of an $SU(2)$ triplet with $Y\ne 0$ would require a higher level
embedding of $SU(2)$, e.g. $SU(2) \subset SU(2) \x SU(2)$, with the $T$ and $\bar{T}$
transforming as $(2,2)$ and the lepton (and Higgs) doublets as $(2,1)$ or $(1,2)$\footnote{Explicit
$Z_3$ constructions have been found with some but not all of these features~\cite{string}.}.
The underlying $SU(2) \x SU(2)$ would only allow off-diagonal trilinear couplings such as
\beq  W \sim 
\lambda^T_{1j}  L_1(2,1) T(2,2) L_j(1,2), \ j = 2,3, \label{embedding} \eeq
yielding
\beq  m_\nu=  \left(
\begin{array}{ccc} 0 & a & b \\ a & 0 & c \\
b& c & 0 \end{array} \right). \label{triptex} \eeq
$c=0$ for the example in (\ref{embedding}), but it could be present
for the embedding $SU(2) \subset SU(2) \x SU(2)\x SU(2)$.
Alternatively, a non-zero but suppressed $c$
 (or the diagonal elements) could be generated by
higher dimensional operators (HDO).

This reasoning provides a stringy motivation to study the neutrino mass matrix
in (\ref{triptex}). In string constructions it is also plausible (but not necessary) to 
assume $|a| = |b|=|c|$  or $|a| = |b| \gg |c|$. 
There are enough
zeroes in (\ref{triptex}) so that one
can take $a,b,c$ real w.l.o.g. by a redefinition of fields. 
Then $ m_\nu= m_{\nu}^{ \dagger}$ with
 $ {\rm  Tr \ } m_\nu = 0$, which implies
$  m_1 + m_2 + m_3 = 0$, where the eigenvalues $m_i$ are real but can be either 
positive or negative. This simple constraint, combined with the observed values
$|\Delta m^2_{\rm  Atm}| \sim 2 \x 10^{-3}$ eV$^2$ and 
$\Delta m^2_\odot \sim 8 \x 10^{-5}$ eV$^2$, leads to the prediction\footnote{There is a 
second solution corresponding to a partially degenerate normal hierarchy for
$|a| \sim |b| \sim |c|$, but this does not lead to realistic mixings.}  
of an inverted hierarchy, $m_i = 0.046, \ -0.045, \ -0.001$ eV.
This corresponds to $|a|\sim |b| \gg |c|$, and approximate bi-maximal mixing,
i.e., $\theta_\odot \sim \theta_{\rm Atm} \sim \pi/4$ for $U_e=I$ ($|a| \ne |b|$
would lead to maximal {\em Solar} mixing and non-maximal atmospheric).
It is convenient to choose the phases of the fields so that $a \sim -b$. 

The limiting case $a=-b$, $c=0$ of (\ref{triptex}) has actually been studied previously by many 
authors~\cite{special}, motivated by bottom-up or other theoretical considerations.
There is a conserved nonstandard lepton number $L_e-L_\mu-L_\tau$,
bimaximal mixing, and an inverted hierarchy with the degenerate pair
forming the variant form of a Dirac neutrino involving only active states (section
\ref{definitions}). The small Solar mass splitting can be induced by turning on a
small $|c|$ or diagonal element, yielding a pseudo Dirac $\nu$.

One cannot simultaneously obtain the observed $\Delta m^2_\odot $
and the observed deviation of $\theta_\odot$ from maximal in this way except
by a fine-tuned cancellation of two rather large corrections. 
As discussed in Section \ref{numixings}, however, the deviation from
maximal could be due to small deviations of $U_e$ from the identity.
In the string context, in particular,
 there is no reason to assume that there are no mixings in $U_e$, so the
model in (\ref{triptex}) is viable. Assuming, for example, that
\beq U_e^\dagger \sim  \left(
\begin{array}{ccc} 1 & -s^e_{12}  & 0 \\ s^e_{12} & 1 & 0  \\
0 & 0 & 1  \end{array} \right), \eeq
with $s^e_{12} $ chosen so that $\theta_\odot  \sim  \frac{\pi}{4} - \frac{s^e_{12}}{\sqrt{2}}
= 0.56^{+0.05}_{-0.04}$, one predicts
$|U_{e3}|^2  \sim    \frac{(s^e_{12})^2}{{2}} \sim (0.023-0.081)$ at 90\% cl, close to the
present upper limit of 0.03; an observable $\beta \beta_{0\nu}$ mass
$m_{\beta \beta} \sim  m_2 ( \cos^2 \theta_\odot - \sin^2 \theta_\odot)
\sim 0.020\  {\rm   eV}$; and a total cosmological mass $\sum |m_i| = 0.092$ eV.
 
\section{Other Implications}
 Let me briefly mention a number of other implications of neutrino mass.
 \begin{itemize}
 \item Lepton flavor nonconservation  (LFV)~\cite{LFV}, e.g., $\mu \ra e \gamma$,
$\mu N \ra eN$, $\mu \ra 3e$.
 Lepton and hadron FCNC are expected at some level in most BSM theories.
 In principle, nonzero neutrino mass and mixings
  violates $L$ flavor, but the effects are negligible except for neutrino oscillations.
However, significant LFV is often generated along with $m_\nu$ in specific models,
e.g. by $\tilde{\nu}$ exchange in supersymmetry.
\item Large magnetic moments are possible~\cite{rsfp}, though the simplest
neutrino mass models yield very small moments. There are rather stringent astrophysical limits.
\item Massive neutrinos may decay, with implications for high energy~\cite{decay1}, 
supernova~\cite{decay2}, solar~\cite{decay3},
and cosmological $\nu$'s~\cite{decay4}.
\item In addition to the possible CP violating phase in the PMNS matrix, which may be
observable in long baseline experiments, Majorana neutrinos allow two additional 
phases. These are in principle observable in $\beta \beta_{0\nu}$, but in practice this is
difficult due to  nuclear uncertainties~\cite{mphase}.
\item Oscillation effects will likely lead to an equilibration of lepton asymmetries
between lepton flavors. This greatly strengthens the limits on asymmetries from BBN~\cite{equil}
unless there was a compensating contribution to the energy density in the early universe~\cite{bbnasym}.
\item High energy $\nu$'s are expected from violent astrophysical events, such as active
galactic nuclei and gamma ray bursts. Measurements of the $\nu_e/\nu_\mu/\nu_\tau$ ratio 
would be very sensitive to oscillations and decays~\cite{henu}. 
$Z$-bursts (the annihilation of an ultra high energy $\nu$ with a relic $\nu$ into a $Z$) could allow
the detection of relic $\nu$'s, but only for an unexpectedly high flux~\cite{zburst}.

\item The most popular model for baryogenesis is leptogenesis~\cite{leptogenesis},
in which the asymmetric decays of the  heavy Majorana neutrino in a seesaw model, 
$N \ra lH \ne N \ra \bar{l} \bar{H}$ can generate a lepton  asymmetry. This $L$ asymmetry is
then partially converted to a $B$ asymmetry ($n_B/n_\gamma \sim 6 \x 10^{-10}$)
by electroweak $B+L$-violating thermal fluctuations  (sphalerons) 
prior to the electroweak phase transition. There are severe constraints on
this mechanism in supersymmetric models because of difficulties for BBN
due to the decays of gravitinos produced after inflation~\cite{moroi}. However, these
can be avoided for some versions of supersymmetry or if the heavy neutrinos
are produced nonthermally. The relevant CP phases are unfortunately not directly
measurable  at low energies, although there may be model-dependent relations to LFV in supersymmetry~\cite{seesawphases}. There are alternative forms
of leptogenesis associated with heavy triplet models~\cite{triplet,tripletlepto}.
There are also viable mechanisms for baryogenesis~\cite{baryogenesis} not related to neutrinos,
such as electroweak baryogenesis (especially in extensions of the MSSM~\cite{bmssm}),  the Affleck-Dine
mechanism, etc.

 \end{itemize}

\section{Conclusions}
\begin{itemize}
\item Nonzero neutrino masses are the first necessary extension of the standard model.
\item  The experimental program has been spectacularly successful.
\item $m_\nu$ may well  be due to GUT or Planck scale physics.
\item There are many possibilities for neutrino mass, both Dirac and Majorana.
In particular, string constructions are unlikely to yield the
standard GUT or left/right motivated seesaw models. One should allow for
the possibilities of small Dirac masses, non-standard or extended seesaw models,
and triplet models, perhaps with an inverted hierarchy.
\end{itemize}


\end{document}